\documentclass[floatfix,twocolumn,aps,prd,showpacs,amsmath,amssymb]{revtex4-1}
\usepackage[latin1]{inputenc}
\usepackage[dvips]{graphicx}
\usepackage{dcolumn}
\usepackage{bm}
\usepackage{dsfont}
\usepackage{color}
\usepackage{ulem} 
\usepackage{url}
\usepackage{amsmath,amssymb}

\newcommand{\PR}[1]{\ensuremath{\left[#1\right]}}
\newcommand{\PC}[1]{\ensuremath{\left(#1\right)}}
\newcommand{\chav}[1]{\ensuremath{\left\{#1\right\}}}

\begin{document}
\title{The influence of a weak magnetic field in the Renormalization-Group functions of (2+1)-dimensional Dirac systems}

\author{Nat\'alia Menezes$^{1}$, Van S\'ergio Alves$^{2}$, C. Morais Smith$^1$ }

\affiliation{$^1$Institute for Theoretical Physics, Center for Extreme Matter and Emergent Phenomena, Utrecht University, Leuvenlaan 4, 3584CE Utrecht, the Netherlands\\
$^2$Faculdade de F\'\i sica, Universidade Federal do Par\'a, Avenida Augusto Correa 01, 66075-110, Bel\'em, Par\'a,  Brazil }

\begin{abstract}
The experimental observation of the renormalization of the Fermi velocity $v_{F}$ as a function of doping has been a landmark for confirming the importance of electronic interactions in graphene. Although the experiments were performed in the presence of a perpendicular magnetic field $B$, the measurements are well described by a renormalization-group (RG) theory that did not include it. Here we clarify this issue, for both massive and massless Dirac systems, and show that for the weak magnetic fields at which the experiments are performed, there is no change in the renormalization-group functions. Our calculations are carried out in the framework of the Pseudo-quantum electrodynamics (PQED) formalism, which accounts for dynamical interactions. We include only the linear dependence in $B$, and solve the problem using two different parametrizations, the Feynman and the Schwinger one. We confirm the results obtained earlier within the RG procedure and show that, within linear order in the magnetic field, the only contribution to the renormalization of the Fermi velocity arises due to interactions. In addition, for gapped systems, we observe a running of the mass parameter.
\end{abstract}

\maketitle

\section{Introduction}
The synthesis of graphene \cite{Novoselov2004}, a two-dimensional material composed of carbon atoms organized in a honeycomb lattice, had a huge impact in condensed-matter physics. Due to the lattice geometry, this material has two inequivalent Dirac points ($K$ and $K'$), each one associated to a valley degree of freedom. In the vicinity of these points, the free electrons exhibit a linear dispersion relation, i.e., $E \propto v_{F}|\textbf{k}|$, where $v_{F}$ is the Fermi velocity, which has a bare value three hundred times smaller than the speed of light.

After graphene, other layered two-dimensional materials with similar properties have been realized, such as silicene \cite{silicene}, stanene \cite{stanene}, germanene \cite{germanene} and transition metal dichalcogenides (TMDCs) \cite{ReviewTMDCs}. Unlike graphene, which has a gapless spectrum, these other layered materials present an intrinsic bandgap. Silicene, stanene and germanene are semiconductors represented by a single-atom species. Instead of carbon atoms, this other class of materials is composed by heavier atoms (e.g., silicon, germanium). When these atoms with larger ionic radius assemble to form honeycomb structures, the lattices are not flat like graphene, but buckled, which leads to the gap in the spectrum. On the other hand, TMDCs consist of layers composed of more than one-atom species. The TMDCs layers are weakly bonded by Van der Waals interactions, which permits their treatment as a two-dimensional system. Chemically, the TMDCs' composition is represented as MX$_{2}$, where $M$ is the transition-metal atom (Mo, W etc.) and $X$ is the chalcogen atom (Se, S or Te). According to the choice of atoms, these layered materials can exhibit a wide range of physical properties, which includes superconducting, magnetic or topological-insulating behavior, for example. The wide bandgap present in monolayer TMDCs is very convenient for electronic applications \cite{ReviewTMDCs}. 

For all these materials, the Fermi velocity is an important parameter that characterizes the system. Therefore, a relevant question in the description of the Dirac electrons in these systems is how the Fermi velocity renormalizes due to interactions. Even before the isolation and characterization of graphene, this question was answered through field-theoretical studies that have predicted the effect of interactions in two-dimensional massless Dirac systems, where the electrons and the photons can live in different dimensions \cite{Marino1993, Maria1994}. Indeed, both in graphene and related gapped 2D systems, the electrons are constrained to move on a plane, while the mediators of the interaction (photons) can propagate in a three-dimensional space. Differently from usual quantum electrodynamics (QED) in (2+1) dimensions, these kind of effective theories generate a Coulomb potential between the electrons proportional to the inverse of the distance, similar to QED in (3+1)D. 

A renormalization-group study of graphene predicted logarithmic corrections to the Fermi velocity as a function of doping (or energy) \cite{Maria1994, Foster2008, Maria1996, Maria2010, Maria2012}, which were later observed in many experiments \cite{Elias2011,Luican2011,Stroscio2012}. In addition, the renormalized $v_{F}$ also depends strongly on the dielectric constant of the medium surrounding the graphene sample.

The experimental confirmation of this renormalization called the attention to the role of interactions in graphene and other 2D condensed-matter systems that can be described by relativistic Dirac electrons. Moreover, since the Fermi velocity is the characteristic velocity of the system, all the physical observables carry this information, and this effect is also seen in indirect measurements, e.g., in the quantum capacitance \cite{Yu2013} and in the spin $g$-factor \cite{Kurganova2011, Stroscio2010}. A theoretical description of the corrections to the $g$-factor due to interactions can only account for the experimental data upon insertion of the renormalized Fermi velocity and dielectric constant as a function of doping \cite{Menezes2016}.

Although theoretical studies have clarified the role of interactions in renormalizing the Fermi velocity, most of the experiments verifying this behavior are performed in the presence of a magnetic field. The remaining question, to be answered theoretically, is then whether the renormalization-group functions are modified or not due to a magnetic field applied perpendicularly to the graphene plane. 

A study of the Schwinger-Dyson equations in the \textit{static} limit in the presence of a magnetic field suggests a renormalization of the Fermi velocity in each of the Landau levels due to electron-electron interactions \cite{Gorbar2012}. On the other hand, the experimental findings are well fitted by a renormalization-group description that ignores the magnetic field. An important issue in this comparison is the intensity of the magnetic field. Although the calculations in Ref.~\cite{Gorbar2012} are made in the ``weak"  field approximation \cite{footnote}, they cannot describe the experiments detecting the renormalization of the Fermi velocity \cite{Elias2011,Luican2011,Stroscio2012} because these experiments are not in the Landau-level, but in the Shubnikov-de Haas regime.

Here, we investigate this problem within the Pseudo-quantum electrodynamics (PQED) framework, which accounts for \textit{dynamical} interactions, using a field-theoretical method. Since PQED is a renormalizable theory, i.e., the coupling constant is dimensionless, we use perturbation theory up to one-loop order to obtain the first correction to the fermionic propagator due to interactions, and under the presence of a weak external magnetic field. We show that in the weak-field approximation, we may separate the electron self-energy in two pieces: one at zero magnetic field, and another with a linear dependence on the field. Focusing only on the $B$-field term, through two different parametrizations, Feynman's and Schwinger's, we compute the contribution due to the magnetic field, which happens to be finite. Within the renormalization group equations, we show that neither the weak magnetic field nor any finite contribution modify the renormalization of the Fermi velocity. In addition, for gapped systems we find that the mass renormalizes and its flow depends on the strength of the interaction.

The paper is divided as follow. In Sec.~II, we introduce the PQED model used in our calculations, and the Feynman rules associated with it, in the presence of an external magnetic field $B$. In Sec.~III, we compute the electron self-energy in the weak-field approximation using two different parametrizations, for both the massive and massless cases. In Sec.~IV, we outline the renormalization-group equations for the model in order to investigate the effect of the weak magnetic field and check the running of the mass parameter. We present the conclusions of our work in Sec.~V. The details of the calculations are given in the appendices.

\section{The model}

\begin{figure}[!htb]
	\centering
		\includegraphics[width=0.4\textwidth]{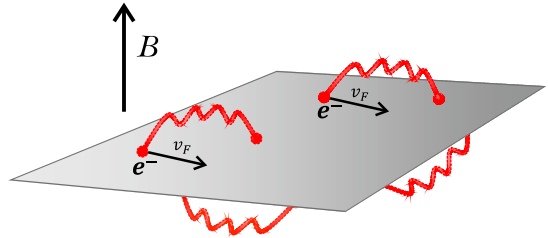}
	\caption{Illustrative picture of the system studied.}
	\label{Systemfig}
\end{figure}

The particular system of our interest is illustrated in Fig.~\ref{Systemfig}. There are electrons propagating with a Fermi velocity $v_{F}$ in a two-dimensional space, under the influence of an external magnetic field applied perpendicularly to it. Moreover, the photons through which the electrons interact are not confined to the plane, and can propagate in a three-dimensional space. 

Mathematically, the dimensional mismatch illustrated above can be described by imposing a constraint in the matter current, and the result is a projected theory called Pseudo-QED \cite{Marino1993}. This effective theory works in (2+1)D, and the term ``pseudo'' originates due to the pseudo-differential operator that now appears in the Maxwell Lagrangian (see Eq.~(\ref{PQEDlagrangian})).

The Pseudo-QED Lagrangian, in the presence of an external magnetic field, is given by
\begin{eqnarray}
\mathcal{L} &=& -\frac{1}{2} \frac{F_{\mu\nu}F^{\mu\nu} }{\sqrt{\Box}}+ \bar{\psi}\PR{i \gamma^{\mu}\PC{\bar{\partial_{\mu}} -eA_{\mu}}-m}\psi , \label{PQEDlagrangian}
\end{eqnarray}
where $\Box=c^{2}\Delta - \partial^{2}/\partial t^{2}$, $\gamma^\mu=(\gamma^0,\beta\gamma^i)$, $\bar{\partial_{\mu}}=(\partial_{0},v_{F}\partial_{i})$, $A_{\mu}=(A_{0}, A_{i})$, $F_{\mu\nu}=\partial_{\mu}A_{\nu}-\partial_{\nu}A_{\mu}$, $m$ is the fermionic mass and the dimensionless parameter $\beta=v_{F}/c$. Now, the minimal coupling is written as a sum of a quantum $A_{\mu}^{(q)}$ and a classical $A_{\mu}^{(e)}$ contributions, i.e., $A_{\mu} = A_{\mu}^{(q)} + A_{\mu}^{(e)}$. The first term is the vector potential associated to the quantized dynamical electromagnetic field, which describes the interaction between the photon and the fermion fields, whereas the second is due to the external magnetic field. In this work, we adopt the Landau gauge $A_{\mu}^{(e)}=(0,0,Bx)$, with $B$ denoting a constant magnetic field that couples minimally to the free-fermion momentum to generate the discrete Landau levels.

The Schwinger's proper-time representation of the fermion propagator in (2+1)D in momentum space $k$ is \cite{Schwinger48} 
\begin{eqnarray}
&&S_{F}(\bar{k})= \int_{0}^{\infty} ds e^{is\PC{k_{0}^{2}+i\eta-m^{2}}-iv_{F}^{2}\textbf{k}^{2}\ell^{2}\tan{(s|eB|)}} \nonumber \\
&&\times\PR{k_{0}\gamma^{0}-v_{F}\textbf{k}\cdot\gamma-m-v_{F}(k^{1}\gamma^{2}-k^{2}\gamma^{1})\tan{(s|eB|)}} \nonumber \\
&&\times \PR{1+\gamma^{1}\gamma^{2}\tan{(s|eB|)}}, \label{SF}
\end{eqnarray}
where ${\bar k}^{\mu}=(k_0,v_F \textbf{k})$ is the electron momentum with $\bar k^2=k_0^2-v_F^2 \textbf{k}^2$, the parameter $s$ is the proper time of the particles while they travel throughout their paths in the Feynman diagram \cite{Schwinger48}, $\eta$ is the \textit{causal} factor, and $\ell = \sqrt{c (|eB|)^{-1}}$ (we assume $\hbar=1$). The $\gamma^{1,2}$ and the $k^{1,2}$ are the spatial components of the $\gamma$-matrices and the momentum, respectively. Here, we neglect finite-density contributions because we are interested in the behavior of the system near the Dirac points. Perturbative calculations taking into account these extra contributions were performed in QED$_{2+1}$ \cite{Khalilov} and QED$_{3+1}$ \cite{PRD88}.

The poles of the fermionic propagator yield the energy dispersion relation $p_{0}=\pm E_{n}=\pm \sqrt{2|eB|n+m^{2}}$, where $n$ is the quantum number associated with the discrete Landau levels \cite{Khalilov}. The photon propagator in the Landau gauge and the interaction vertex are defined, respectively, as 
\begin{eqnarray}
\Delta_{\mu\nu} (k) &=& \frac{-icg_{\mu\nu}}{2\varepsilon\sqrt{k^2}},  \\
\Gamma^{\mu}_{0}&=&-ie\PC{ \gamma^{0}, \beta\gamma^{j}}, 
\end{eqnarray}
where $g_{\mu\nu}=(+,-,-)$, $\varepsilon$ is the dielectric constant, and the photon momentum is $k_{\mu}=(k_0,c\textbf{k})$ with $k^2=k_0^2-c^2 \textbf{k}^2$.

\section{Electron self-energy}

\begin{figure}[!htb]
	\centering
		\includegraphics[width=0.25\textwidth]{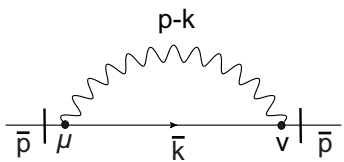}
	\caption{Electron self-energy. The bar symbol on top of the momenta is used to denote the electron momentum, which enters with the Fermi velocity $v_{F}$, contrarily to the photon propagator, which occurs with the speed of light.}
	\label{FigSelfE}
\end{figure}

The electron self-energy $\Sigma$, represented by the Feynman diagram given in Fig.~\ref{FigSelfE}, carries the information about the propagation of the electron under the effect of interactions. Therefore, to investigate the possible renormalization of the parameters contained in the Dirac Lagragian, i.e., the Fermi velocity, the electron mass and the fermionic field itself, one needs to calculate $\Sigma$. First, we will analyze the zero-mass case, and then discuss what changes in the presence of the fermionic mass. 

\subsection{The zero-mass case}

In one-loop order, the diagram represented in Fig.~\ref{FigSelfE} reads
\begin{eqnarray}
\Sigma (\bar{p})&=& i\int \frac{d^{3}k}{(2\pi)^{3}}\Gamma^{\mu}_{0}S_{F}(\bar{k})\Gamma^{\nu}_{0}\Delta_{\mu\nu}(p-k)\nonumber \\
&=& -\frac{(1-2\beta^{2})ce^{2}}{2\varepsilon}\int_{0}^{\infty} ds \int\frac{d^{3}k}{(2\pi)^{3}} \exp\Big[is\PC{k_{0}^{2}+i\eta}\nonumber \\
&-&iv_{F}^{2}\textbf{k}^{2}\ell^{2}\tan{(s|eB|)}\Big] \frac{k_{0}\gamma^{0}a_{1}(B) - v_{F}\textbf{k}\cdot\gamma a_{2}(B)}{\sqrt{(k_{0}-p_{0})^2-c^{2}(\textbf{k}-\textbf{p})^2}},\nonumber \\ \label{Self-energy}
\end{eqnarray}
where $a_{1}(B)= 1+\gamma^{1}\gamma^{2}\tan{\PC{s|eB|}}$, and $a_{2}(B)= 1+\tan^{2}{\PC{s|eB|}}$ (for more details of the calculations see Appendix A).
Using Schwinger's parametrization,
\begin{eqnarray}
\frac{1}{A^{z}}&=&\frac{(-i)^{z}}{\Gamma(z)}\int_{0}^{\infty} d\xi \xi^{z-1}e^{i\xi A},\label{Parameter-Sch}
\end{eqnarray}
we may rewrite the self-energy as
\begin{eqnarray}
&&\Sigma (\bar{p})=-\frac{(1-2\beta^{2})ce^{2}}{2(i\pi)^{1/2}\varepsilon}\int_{0}^{\infty} \frac{d\xi}{\xi^{1/2}}\int_{0}^{\infty} ds\int \frac{d^{3}k}{(2\pi)^{3}} \Big[k_{0}\gamma^{0}a_{1} \nonumber \\
&&+ v_{F}\textbf{k}\cdot\gamma a_{2} \Big]e^{i(s+\xi)\PC{k_{0}-\frac{\xi p_{0}}{s+\xi}}^{2}-iD\PC{\textbf{k}-\frac{\xi c^{2}\textbf{p}}{D}}^{2}-\Delta} , \label{dis}
\end{eqnarray}
where 
\begin{eqnarray}
\Delta (p_{0},\textbf{p}) &\equiv& -i\xi p_{0}^{2}\PC{1-\frac{\xi}{s+\xi}}+i\xi c^{2}\textbf{p}^{2}\PC{1-\frac{\xi c^{2}}{D}}, \nonumber\\
D(B) &=& v_{F}^{2}\ell^{2}\tan{(s|eB|)}+\xi c^{2}. \nonumber
\end{eqnarray}
Shifting the variables in Eq.~(\ref{dis}) as $ k_{0} \rightarrow k_{0}+ \xi p_{0}/(s+\xi)$, $\textbf{k} \rightarrow \textbf{k}+ \xi c^{2}\textbf{p}/D$, and then evaluating the integrals over $\textbf{k}$ and $k_{0}$ (more details in Appendix A), we obtain
\begin{eqnarray}
\Sigma (\bar{p})&=& -\frac{i(1-2\beta^{2}) \alpha \beta}{4\pi} \PC{p_{0}\gamma^{0}I_{1} + v_{F}\textbf{p}\cdot\gamma I_{2}}, \label{struc}
\end{eqnarray}
where $\alpha=e^{2}/4\pi\varepsilon v_{F}$ and the $I_{i}$'s are the following parametric integrals: 
\begin{eqnarray}
&&I_{1}=\int_{0}^{\infty} d\xi \int_{0}^{\infty} ds \frac{\xi^{1/2} a_{1}(B)}{(s+\xi)^{3/2}\PR{\beta^{2}\ell^{2}\tan{(s|eB|)}+\xi }} \times \nonumber \\
&& \exp\chav{i\frac{s\xi p_{0}^{2}}{s+\xi}-i\xi v_{F}^{2}\beta^{-2}\textbf{p}^{2}\PC{1-\frac{\xi }{\beta^{2}\ell^{2}\tan{(s|eB|)}+\xi }}}, \nonumber \\
&&I_{2}= \int_{0}^{\infty} d\xi\int_{0}^{\infty} ds \frac{\xi^{1/2} a_{2}(B)}{(s+\xi)^{1/2}\PR{\beta^{2}\ell^{2}\tan{(s|eB|)}+\xi }^{2}} \times \nonumber \\
&& \exp\chav{i\frac{s\xi p_{0}^{2}}{s+\xi}-i\xi v_{F}^{2}\beta^{-2}\textbf{p}^{2}\PC{1-\frac{\xi }{\beta^{2}\ell^{2}\tan{(s|eB|)}+\xi }}} \nonumber .
\end{eqnarray}

Until now, we considered the full Landau-levels contribution to the one loop self-energy. Nonetheless, to solve analytically the parametric integrals and proceed with a more intuitive expression for the self-energy, it is necessary to examine some approximations. The first useful one is to consider only terms up to linear order in $\beta=v_{F}/c$. Since linear terms in $\beta$ are already of order of $1/300$, second- or higher-order terms would generate negligible contributions that can be discarded. Hence, we have
\begin{eqnarray}
 I_{1}&\approx& \int_{0}^{\infty} d\xi \int_{0}^{\infty} ds \frac{a_{1} \exp\PR{i\frac{s\xi}{\xi+s}p_{0}^{2}-i\frac{v_{F}^{2}\textbf{p}^{2}\tan{\PC{s|eB|}}}{|eB|}} }{\sqrt{\xi}(s+\xi)^{3/2}},\nonumber \\
 I_{2}&\approx& \int_{0}^{\infty} d\xi\int_{0}^{\infty} ds \frac{a_{2} \exp\PR{i\frac{s\xi}{\xi+s}p_{0}^{2}-i\frac{v_{F}^{2}\textbf{p}^{2}\tan{\PC{s|eB|}}}{|eB|}} }{\xi^{3/2}\sqrt{s+\xi}}.\nonumber
\end{eqnarray}

\subsection{Weak magnetic field approximation}

The second approximation concerns the magnetic field. In the weak-field expansion, one may retain only terms which are up to linear order in the $B$-field. In this case, the parametric integrals become
\begin{eqnarray}
I_{1}&\approx& \int_{0}^{\infty} d\xi \int_{0}^{\infty} ds \frac{e^{-\frac{is}{s+\xi}\PR{(s+\xi)v_{F}^{2}\textbf{p}^{2}-p_{0}^{2}\xi}}\PC{1+\gamma^{1}\gamma^{2}s|eB|}}{\sqrt{\xi}(s+\xi)^{3/2}},\nonumber \\ \label{i1}\\
I_{2}&\approx&  \int_{0}^{\infty} d\xi \int_{0}^{\infty} ds \frac{e^{i\frac{s}{s+\xi}\PR{v_{F}^{2}\textbf{p}^{2}(s+\xi)-p_{0}^{2}\xi}}}{\xi^{3/2}\sqrt{s+\xi}}.   \label{i2}
\end{eqnarray}
We observe in Eqs.~(\ref{i1}) and (\ref{i2}) that the linear in $B$ term gives only an extra contribution to the $p_{0}$ component because $I_{2}$ does not depend on $B$ [see also Eq.~(\ref{struc})]. The remaining integrals are just the effect of interactions, as we expect for zero magnetic field. The same result can be obtained if one starts with the fermionic propagator already in the weak-field approximation \cite{PRD88}, i.e., 
\begin{eqnarray}
S_{F}(\bar{k}) &=& \int_{0}^{\infty} ds \exp\PR{is\PC{k_{0}^{2}-v_{F}^{2}\textbf{k}^{2}}}(k_{0}\gamma^{0} -v_{F}\textbf{k}\cdot\gamma \nonumber \\
&+& |eB| s k_{0}\gamma^{0}\gamma^{1}\gamma^{2} + \ldots). \label{SFapprox}
\end{eqnarray}
Therefore, within these approximations, the additional contribution to the electron self-energy due to the magnetic field can be computed separately. In other words, 
$$\Sigma (\bar{p}) = \Sigma^{(0)}(\bar{p})+\Sigma^{(1)}(\bar{p})+\ldots,$$
where $\Sigma^{(0)}$ is the self-energy in the absence of magnetic field, and the expansion follows with the dependence on the $B$-field, as for the propagator in Eq.~(\ref{SFapprox}).

Now, starting from the propagator in the weak-field approximation, and performing the integrals in two different parametrizations in order to double check our results, we find (see Appendix A for details)

{\bf 1. Feynman parametrization}
\begin{eqnarray}
-i\Sigma^{(1)}(\bar{p}) &=& \frac{ i\alpha \beta |eB|}{4\pi}\int_{0}^{1} dx   \frac{p_{0}\gamma^{0}\gamma^{1}\gamma^{2}}{\sqrt{1-x}\PR{v_{F}^{2}\textbf{p}^{2}-p_{0}^{2}(1-x)}} \nonumber \\
 &=& -\frac{ i\alpha \beta |eB|}{2\pi} \frac{\sin^{-1}\PC{\sqrt{\frac{p_{0}^{2}}{v_{F}^{2}\textbf{p}^{2}-p_{0}^{2}}}}}{v_{F}|\textbf{p}|p_{0}}\ p_{0}\gamma^{0}\gamma^{1}\gamma^{2},\qquad \label{result1}
\end{eqnarray}

{\bf 2. Schwinger parametrization}
\begin{eqnarray}
-i\Sigma^{(1)}(\bar{p}) &=& \frac{ \alpha\beta|eB|}{4\pi} \int_{0}^{\infty} d\xi \int_{0}^{\infty}  ds\ \frac{  s e^{is\PR{p_{0}^{2}\PC{\frac{\xi}{s+\xi}}-v_{F}^{2}\textbf{p}^{2}}}}{(s+\xi)^{3/2}\xi^{1/2}} \nonumber\\
&\times& p_{0}\gamma^{0}\gamma^{1}\gamma^{2} \nonumber \\
&=&-\frac{ i\alpha \beta |eB|}{2\pi  }\frac{ \coth^{-1}\PC{\frac{v_{F}|\textbf{p}|}{p_{0}}}}{v_{F}|\textbf{p}|p_{0}}p_{0}\gamma^{0}\gamma^{1}\gamma^{2},  \label{result2} 
\end{eqnarray}
where Eqs.~(\ref{result1}) and (\ref{result2}) are equivalent. Although the results obtained for the two parametrizations may seem different at first glance, below we plot both trigonometric functions together to show their qualitative behavior, and illustrate that the result is indeed independent of the parametrization scheme in the regime of validity of the theory. 
\begin{figure}[!htb]
\centering
		\includegraphics[width=0.4\textwidth]{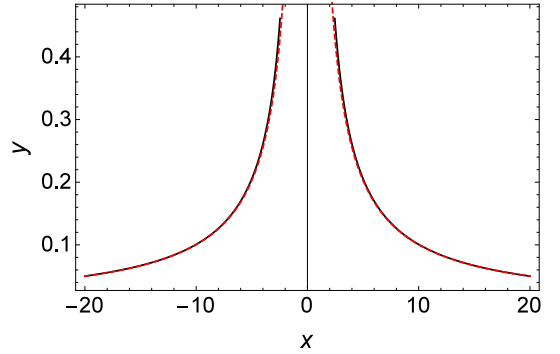}
	\caption{Qualitative comparison between the trigonometric functions in Eqs.~(\ref{result1}) and (\ref{result2}) to show their equivalence. The black solid line represents the inverse of the sine function, and the red dashed line represents the inverse of the cotangent hyperbolic function.}
	\label{Comparison}
\end{figure}

In the $y$-axis in Fig.~\ref{Comparison}, we represent 
$$y(x)=\sin^{-1}\PR{\PC{x^{2}-1}^{-1/2}}$$
with a black solid line, and 
$$y(x)=\coth^{-1}(x)$$ 
with a red dashed line, for a given value of $x=v_{F}|\textbf{p}|/|p_{0}|$. Both trigonometric functions are only valid for ${\rm Re}\PR{|x|} \geq 1$.

These results show that in linear order the magnetic field gives a finite contribution to the electron self-energy.
Although this result suggests that the magnetic field will not modify the flow of the Fermi velocity, in the next section we explicitly calculate the renormalization-group equations to show that this is indeed the case. 

\subsection{The fermionic mass contribution}
Now, we will examine what effectively happens in the study of the self-energy for the massive case. The expansion of the propagator given in Eq.~(\ref{SF}) up to linear order in the magnetic field yields
\begin{equation}
S^{(0)}(\bar{k}) = i\frac{k_{0}\gamma^{0}-v_{F}\textbf{k}\cdot\gamma+m}{k_{0}^{2}-v_{F}^{2}\textbf{k}^{2}-m^{2}},\label{s01}
\end{equation}
and
\begin{equation}
S^{(1)}(\bar{k}) = -eB \frac{k_{0}\gamma^{0}+m}{[k_{0}^{2}-v_{F}^{2}\textbf{k}^{2}-m^{2}]^{2}}\gamma^{1}\gamma^{2}.\label{s11}
\end{equation}

As we have seen already for the massless case, the linear contributions on the magnetic field appeared to be finite and do not affect the renormalization group functions. Therefore, here we will focus on the mass term of Eq.~(\ref{s01}) because this will give us the divergent contribution that will affect the mass renormalization.

Following a standart procedure, we find
\begin{equation}
-i\Sigma^{(0)} (m) = -\frac{\alpha\beta}{2\pi \epsilon}(1+2\beta^{2})mI_{3}, \label{massterm}
\end{equation}
where $\epsilon\rightarrow 0$ and
\begin{equation}
I_{3} = \frac{1}{c^{2}}\int_{0}^{1} dx \frac{x^{-1/2}}{[\beta^{2}(x-1)-x]}. \label{I3}
\end{equation}

\section{Renormalization-Group study}
In order to use the renormalization-group functions, first we need to define the expression for the inverse of the free-fermion propagator in the presence of the magnetic field. This turns out not to be a problem because the information about the $B$-field is contained within the Schwinger's phase factor \cite{Schwinger48} and the inverse of the propagator happens to be the same as in the case of zero $B$-field \cite{PRD88,PRD88-2}. Based on this statement, we can start from the propagator as in Eq.~(\ref{SF}), without any approximations, and obtain an expression for the self-energy with all the possible contributions coming from the magnetic field. Hence, the case of a weak magnetic field would only be considered in the approximation for the self-energy.

The renormalization-group equation is given by
\begin{eqnarray}
\left( \mu \frac{\partial}{\partial \mu} +  \beta_{e}\frac{\partial}{\partial e} + \beta_{v_{F}}\frac{\partial}{\partial v_{F}} + \beta_{c}\frac{\partial}{\partial c}+ \gamma_{m} m\frac{\partial}{\partial m} \right. \nonumber \\ \left.- N_{F}\gamma_{\psi}- N_{A}\gamma_{A}\right) \Gamma^{(N_{F},N_{A})}=0, \label{RGE0}
\end{eqnarray}
where $\Gamma^{(N_{F},N_{A})}$ represent the vertex functions, with $N_{F}$ and $N_{A}$ the number of fermion and photon external lines, respectively, in the Feynman diagrams. The functions $\gamma_{\psi}$ and $\gamma_{A}$ are the respective anomalous dimension of the fermion and photon fields, $m\gamma_{m} = \mu\partial m/\partial \mu$ is a dimensionless function for the mass, and $\beta_{i}$ ($i=e,v_{F},c$) are the beta-functions associated to the parameters of the Pseudo-QED Lagrangian. We use dimensional regularization to obtain the vertex functions in Eq.~(\ref{RGE0}). 

In the case of the fermion two-point function, we have
\begin{eqnarray}
\left( \mu \frac{\partial}{\partial \mu} +  \beta_{e}\frac{\partial}{\partial e} + \beta_{v_{F}}\frac{\partial}{\partial v_{F}} + \beta_{c}\frac{\partial}{\partial c}+\gamma_{m} m\frac{\partial}{\partial m}\nonumber \right.\\
\left.- 2\gamma_{\psi} \right) \Gamma^{(2,0)}=0, \label{RGE}
\end{eqnarray}
with
\begin{eqnarray}
\Gamma^{(2,0)} = -i\PC{\gamma^{0}p_{0}+v_{F}\gamma^{i}p_{i}+m} - i\Sigma.\label{1Gammma20}
\end{eqnarray}
Here, we write the self-energy $\Sigma$ in a general form, where all the possible contributions of an external magnetic field could be included. 

According to our approximation $\Sigma \approx \Sigma^{(0)} + \Sigma^{(1)}$, the self-energy can then be written as
\begin{eqnarray}
- i\Sigma =  e^{2}\PC{{\rm finite}^{(2,0)}+\ln\mu {\rm Res}^{(2,0)}} + e^{3}f(B) ,\label{Gammma20} \qquad  
\end{eqnarray}
where we divide the zero magnetic field part into a finite and a divergent contribution, with 
\begin{eqnarray}{\rm Res}^{(2,0)} = A_{1}\gamma^{0}p_{0}+ A_{2}\gamma^{i}p_{i}+A_{3}m,\label{Residue}\end{eqnarray}
representing the pole term proportional to $1/\epsilon$. For the RG purposes, here, the explicit form of the finite$^{(2,0)}$ contribution is irrelevant. 
The coefficients $A_{1}$ and $A_{2}$ are determined using Eq.~(\ref{struc}) for $B=0$, whereas $A_{3}$ is determined from Eq.~(\ref{massterm}), and the function $f(B)$ is the finite result obtained from Eq.~(\ref{result1}) or (\ref{result2}). 

\subsubsection{Velocity renormalization}

Now, expanding each one of the parameters in Eq.~(\ref{RGE}) in terms of the coupling constant $e$, e.g.,
\begin{eqnarray} 
\beta_{v_{F}}&=& \beta_{v_{F}}^{(1)} e +  \beta_{v_{F}}^{(2)} e^{2} + \beta_{v_{F}}^{(3)} e^{3} + \ldots , \nonumber
\end{eqnarray}
going up to third order, and applying Eq.~(\ref{RGE}), we find that $\gamma^{(1)}_{\psi}=\beta_{v_F}^{(1)}=0$. Moreover, performing the same analysis for the other two vertex functions, $\Gamma^{(2,1)}$ and $\Gamma^{(0,2)}$, we find that $\gamma^{(1)}_{A}=\beta_{c}^{(1)}=\beta_{e}^{(1)}=0$ (for more details of the calculations see Appendix B). In second order in the coupling constant, for $\beta_{v_F}^{(2)}$, we obtain the well-known renormalization of the Fermi velocity solely due to interaction effects \cite{Maria1994}. This is expected because the magnetic-field term enters in Eq.~(\ref{Gammma20}) as $e^{3}$, hence, the only possible contribution should be seen in this order of the coupling constant. At third order in $e$, we observe that the corrections to $\beta_{v_{F}}^{(3)}$, depending on the finite part of the self-energy, would appear for $\beta_{e}^{(2)}\neq 0$. However, $\beta_{e}^{(2)} \propto \gamma_{A}^{(1)}$, and as the photon self-energy has no divergences in one-loop order, using dimensional regularization, its anomalous dimension is null ($\gamma_{A}^{(1)}=\gamma_{A}^{(2)}=0$). Thus, $\beta_{v_{F}}^{(3)}=0$, and no additional renormalization term is generated due to the presence of an external magnetic field. 

The fact that only the $B=0$ term in Eq.~(\ref{SF}) contributes to the renormalization of the parameters in the Lagrangian (\ref{PQEDlagrangian}) may suggest that the distinction between weak- or strong-field limit is irrelevant. However, the weak- or strong-field case is determined by the comparison between the two length scales in the theory, namely the magnetic length $\ell_B \propto B^{-1/2}$ and doping $\ell_n \propto n^{-1/2}$. The renormalization-group flow is suppressed and stops at the largest length (or smallest energy) scale; hence, at the critical point ($n\approx 0$) the doping energy is the one that determines the cutoff. 

\subsubsection{The running mass}

The second-order expansion in the coupling constant yields to the mass function 
\begin{eqnarray}
\gamma_{m}^{(2)} &=& - i(A_{3}+ A_{1})\nonumber\\
&=& - \frac{e^{2}}{8\pi^{2}c\varepsilon}\int_{0}^{1}dx \frac{x^{1/2}(1-2\beta^2)+x^{-1/2}(1+2\beta^2)}{\beta^{2}(x-1)-x}\nonumber \\
&=&  - \frac{\alpha}{2\pi} F(\beta),\label{gammam}
\end{eqnarray}
where
\begin{eqnarray}
F(\beta)&=& 2\frac{(1-\beta^{2}+\beta^{4}){\rm ArcTan}\PR{(-1+\beta^{-2})^{1/2}}}{(-1+\beta^{-2})^{1/2}(-\beta+\beta^{3})}\nonumber \\
&+&2\frac{(\beta^{2}-2\beta^{4})}{(-\beta+\beta^{3})} .\nonumber
\end{eqnarray}

Now, calculating Eq.~(\ref{gammam}) on the fixed point of the theory ($\beta=1$), we obtain
\begin{eqnarray}
\lim_{\beta\rightarrow 1}\gamma_{m}^{(2)} = \frac{5\alpha}{3\pi}. \label{GM}
\end{eqnarray}

The mass parameter runs as
\begin{eqnarray}
\frac{\partial \ln m(\mu)}{\partial \ln (\mu/\mu_0)} = \gamma_{m}^{(2)} (\beta), \label{mruns} 
\end{eqnarray}
and integrating Eq.~(\ref{mruns}), we obtain
\begin{eqnarray}
m(\mu) = m_{0} \PC{\frac{\mu}{\mu_0}}^{\gamma_m^{(2)}} \approx m_{0} \PC{\frac{\mu}{\mu_0}}^{5\alpha/3\pi},
\end{eqnarray}
with $m_0=m(\mu_0)$. We see from Eq.~(\ref{GM}) that $\gamma_{m}^{(2)}$ has a positive sign and depends on $\alpha$. 

These are the two main results of this paper: first, the magnetic field does not renormalize any of the parameters of the Lagrangian (\ref{PQEDlagrangian}), and second, the interaction defines how fast the mass parameter runs. Furthermore, as expected, the mass parameter cures infrared divergences that may arise due to the $B$-field expansion.

\section{Conclusions}
Motivated by the fact that most of the experiments of the Fermi velocity renormalization in graphene are performed in the presence of a weak external magnetic field \cite{Elias2011,Luican2011,Stroscio2012}, whereas the field-theoretical models either ignore the latter \cite{Maria1994} or study the problem in the (strong field) Landau-level regime \cite{Gorbar2012}, we decided to revise the topic.

Our starting point is the Pseudo-QED formalism, which accounts for dynamical interactions, under the presence of a weak perpendicular magnetic field. The magnetic field contribution to the self-energy was obtained using two different but equivalent parametrization schemes. The analysis of the renormalization group shows that a weak magnetic field has no additional effect in the renormalization of the Fermi velocity, within linear order in $B$. In this particular theory, because the photon field has null anomalous dimension, up to third-order in the coupling constant $e$, no finite contributions coming from the electron self-energy can modify this renormalization. Hence, in this approximation, it is sufficient to consider only the effect of interactions to observe how the velocity changes with respect to the energy scale of the theory.

It has been observed in Ref.~\cite{Yu2013}, through measurements of quantum capacitance, that the Fermi velocity displays the same indistinguishable logarithmic renormalized behavior as a function of doping both in the absence or in the presence of a weak magnetic field. Our results confirm that, from a theoretical perspective, this should be indeed the case. 

A simple analysis of the perturbation theory shows that our results hold also for high-order loops due to the fact that the theory is renormalizable. Therefore, in the weak-field expansion, any contribution depending on the magnetic field $B$ would generate additional finite terms to the electron self-energy, which do not change the renormalization-group functions. This result does not depend on the massive or massless nature of the system.

In massive systems, however, we obtain a renormalization of the mass parameter, the flow of which depends on the strength of the interaction $\alpha$. This renormalization effect is solely due to the electron-electron interaction.

Even though the weak magnetic field has no effect in the renormalization-group functions, finite temperatures could affect this renormalization \cite{RG-T1,RG-T2}. In addition, for stronger magnetic fields, it was shown theoretically using the Schwinger-Dyson equations that within the static approximation the interactions renormalize the Fermi velocity with a factor that depends on the Landau-level index \cite{Gorbar2012}. The generalization of this theory to the dynamical case and stronger magnetic fields, however, remains to be done. We hope that our results will stimulate measurements of the renormalization of the Fermi velocity in massive Dirac systems, analogously to experiments performed in graphene. 

\acknowledgments
This work was supported by the CNPq through the Brazilian government project Science Without Borders. We are grateful to Eduardo C. Marino and Vladimir Juricic for fruitful discussions.

\section*{Appendix A: Details of the Self-energy calculations}

\subsection*{Fermionic propagator}
Before introducing the fermionic propagator of Eq.~(\ref{SF}) with $m=0$ in the expression for the self-energy, as shown in Eq.~(\ref{Self-energy}), we combine the $\gamma$-matrices in a compact way, 
\begin{eqnarray}
&& N (\bar{k}) = \PR{k_{0}\gamma^{0}-v_{F}\textbf{k}\cdot\gamma-v_{F}(k^{1}\gamma^{2}-k^{2}\gamma^{1})\tan{(s|eB|)}} \nonumber \\
&\times& \PR{1+\gamma^{1}\gamma^{2}\tan{(s|eB|)}}, \nonumber \\
&=& k_{0}\gamma^{0} \PR{1+\gamma^{1}\gamma^{2}\tan{(s|eB|)}} -  v_{F}\textbf{k}\cdot\gamma\PR{1+\gamma^{1}\gamma^{2}\tan{(s|eB|)}} \nonumber \\
&-&v_{F}(k^{1}\gamma^{2}-k^{2}\gamma^{1})\tan{(s|eB|)}\PR{1+\gamma^{1}\gamma^{2}\tan{(s|eB|)}}, \nonumber \\
&=&  k_{0}\gamma^{0} a_{1}(B) - v_{F}\textbf{k}\cdot\gamma - v_{F}(k^{1}\gamma^{2}-k^{2}\gamma^{1})\tan{(s|eB|)}\nonumber \\
&-&v_{F}(\gamma^{1}k^{1}+\gamma^{2}k^{2})\gamma^{1}\gamma^{2}\tan{(s|eB|)}- v_{F}(k^{1}\gamma^{2}-k^{2}\gamma^{1})\gamma^{1}\gamma^{2}\nonumber \\
&\times&\tan^{2}{(s|eB|)},\nonumber \\
&=& k_{0}\gamma^{0} a_{1}(B) - v_{F}\textbf{k}\cdot\gamma a_{2}(B),\nonumber
\end{eqnarray}
where we use that $\gamma^{1}\gamma^{2}=-\gamma^{2}\gamma^{1}$, $(\gamma^{i})=-1$, and $N (\bar{k})$ is the term that multiplies the exponential in the integrand of Eq.~(\ref{SF}), i.e.,
\begin{eqnarray}
S_{F}(\bar{k})&=& \int_{0}^{\infty} ds N (\bar{k}) \exp\PR{is\PC{k_{0}^{2}+i\epsilon}-iv_{F}^{2}\textbf{k}^{2}\ell^{2}\tan{(s|eB|)}}  \nonumber.
\end{eqnarray}

\subsection*{Integrals over the loop-momentum $k$}

The integrals over $k$ in Sec.~III, after the shift of the variables as
$$k_{0} \rightarrow k_{0} + \frac{\xi p_{0}}{s+\xi}, \quad {\rm and} \quad \textbf{k} \rightarrow \textbf{k} + \frac{\xi c^{2} \textbf{p}}{D},$$
are given by
\begin{eqnarray}
\int_{-\infty}^{\infty}dk_{0}\PC{C_{1}\gamma^{0}k_{0}+C_{2}}e^{i(s+\xi)k_{0}^{2}} = \frac{\pi^{1/2}C_{2}}{(-i)^{1/2}(s+\xi)}, \nonumber
\end{eqnarray}
and
\begin{eqnarray}
&&\int_{-\infty}^{\infty}\int_{-\infty}^{\infty}dk_{1}dk_{2}\PC{C_{3}+C_{4}\gamma^{1}k^{1}+C_{5}\gamma^{2}k^{2}}e^{-iD(k_{1}^{2}+k_{2}^{2})} \nonumber \\
&&= \frac{i\pi C_{3}}{D}, \nonumber
\end{eqnarray}
where
\begin{eqnarray}
C_{1}=a_{1}(B), \qquad C_{4}= C_{5}= v_{F}a_{2}(B),\nonumber\\
C_{3}=\gamma^{0}p_{0}\frac{\xi}{\xi+s}a_{1}(B) + v_{F}\textbf{p}\cdot\gamma \frac{\xi c^{2}}{D}a_{2}(B),\nonumber\\
C_{2}= C_{3} + v_{F}\textbf{k}\cdot\gamma a_{2}(B).\nonumber
\end{eqnarray}

\subsection*{Weak-field limit calculations}

In the weak-field approximation, after integrating the linear contribution in the magnetic field in Eq.~(\ref{SFapprox}), we obtain 
\begin{equation}
S^{(1)}_{F} (\bar k) = -eB\frac{k_{0}\gamma^{0}\gamma^{1}\gamma^{2}}{\PC{k_{0}^{2}-v_{F}^{2}\textbf{k}^{2}}^{2}}\label{SF1},
\end{equation}
and the $B$-field term in the self-energy reads
\begin{eqnarray}
-i\Sigma^{(1)}(\bar{p}) &=& \int \frac{d^{D}k}{(2\pi)^{D}}\Gamma_{0}^{\mu}S_{F}^{(1)}(\bar{p}-\bar{k})\Gamma_{0}^{\nu}\Delta_{\mu\nu}(k) \nonumber \\
&=& -\frac{ice^{2}|eB|}{2\varepsilon}\int\frac{d^{3}k}{(2\pi)^{3}}\frac{(1+2\beta^{2})(p_{0}-k_{0})\gamma^{0}\gamma^{1}\gamma^{2}}{[(p_{0}-k_{0})^{2}-v_{F}^{2}(\textbf{p}-\textbf{k})^{2}]^{2}} \nonumber \\
&\times& (k_{0}^{2}-c^{2}\textbf{k}^{2})^{-1/2}, \label{Sigma1}
\end{eqnarray}
where we used the properties of the $\gamma$-matrices, e.g., $\chav{\gamma^{\mu},\gamma^{\nu}}=2g^{\mu\nu}$, and  $d^{3}k=dk_{0}d^{2}\textbf{k}$. 

Now, to calculate the integrals over the loop momentum $k$ in Eq.~(\ref{Sigma1}), we define which one of the two parametrizations (Feynman's or Schwinger's) will be used. Here, we use Schwinger's parameterization as in Eq.~(\ref{Parameter-Sch}). Nevertheless, if one chooses to use Feynman's parameters, like
$$\frac{1}{D_{1}^{2}D_{2}^{1/2}}=\frac{3}{4}\int_{0}^{1}dx \frac{x(1-x)^{-1/2}}{[D_{1}x+(1-x)D_{2}]^{5/2}},$$
the same result is obtained. As we have shown in Sec.~III, the result should not depend on this choice.

Hence, plugging Schwinger's parameters in Eq.~(\ref{Sigma1}), we find
\begin{eqnarray}
&&-i\Sigma^{(1)}(\bar{p}) = \frac{i^{5/2}ce^{2}|eB|(1+2\beta^{2})}{2\varepsilon\pi^{1/2}(2\pi)^{3}}\int_{0}^{\infty}d\xi \xi \int_{0}^{\infty}d\eta\eta^{-1/2}\nonumber \\
&\times&\int d^{3}ke^{i\PR{\xi\PC{p_0-k_0}^{2}-\xi v_{F}^{2}\PC{\textbf{p}-\textbf{k}}^{2}+\eta(k_{0}^{2}-c^{2}\textbf{k})}}(p_{0}-k_{0})\gamma^{0}\gamma^{1}\gamma^{2}. \nonumber 
\end{eqnarray}
The integrals over $k$ are \textit{Gaussian}, and to solve them we first introduce a regulator $\Lambda^{2}$ to avoid high-energy momentum contributions, e.g., $\exp\PC{-k^{2}\Lambda^{-2}}$. Then, we combine separately the terms proportional to $k_{0}$ and $\textbf{k}$ to complete the square for each of them as in Eq.~(\ref{dis}). The integrals over $k$ yield
\begin{eqnarray}
I_{k} &=& \int dk_{0} e^{i\PC{\xi+\eta+i\Lambda^{-2}}k_{0}^{2}} \int d^2\textbf{k} e^{-i\PC{v_{F}^{2}\xi+c^{2}\eta+i\Lambda^{-2}}\textbf{k}^{2}} \nonumber \\
&=& \frac{\pi^{3/2}}{i^{5/2}c^{2}(\eta+\xi)^{1/2}(\eta+\beta^{2}\xi)},
\end{eqnarray}
where the limit of $\Lambda \rightarrow \infty$ was taken after the integration.

Therefore, 
\begin{eqnarray}
&&-i\Sigma^{(1)}(\bar{p}) = \frac{e^{2}|eB|(1+2\beta^{2})p_{0}\gamma^{0}\gamma^{1}\gamma^{2}}{16\pi^{2}\varepsilon c}  \nonumber \\
&\times& \int_{0}^{\infty}d\xi \int_{0}^{\infty}d\eta\frac{\xi\eta^{1/2}e^{i\PR{ p_{0}^{2}\PC{\frac{\eta\xi}{\eta+\xi}}-v_{F}^{2}\textbf{p}^{2}\PC{\frac{\eta\xi}{\eta+\beta^{2}\xi}}   }}}{(\eta+\xi)^{3/2}(\eta+\beta^{2}\xi)}, \nonumber 
\end{eqnarray}
and for $\beta^{2}\rightarrow 0$, we obtain the result given in Eq.~(\ref{result2}).

\section*{Appendix B: RG Calculations}
In this appendix, we show more details of the calculations concerning the renormalization-group equations. As usual, the scaling parameter $\mu$ is introduced through $\mu^{\epsilon/2}$, where $\epsilon$ will be taken to zero in the end. Hence, applying Eq.~(\ref{1Gammma20}) in Eq.~(\ref{RGE}), with $\Sigma$ given by Eq.~(\ref{Gammma20}), we find the following partial derivatives
\begin{eqnarray}
&& \mu \frac{\partial \Gamma^{(2,0)}}{\partial \mu}=e^{2}{\rm Res}^{(2,0)}, \nonumber \\
&& \beta_{e}\frac{\partial \Gamma^{(2,0)}}{\partial e} =  \beta_{e} \PR{2e\PC{\tilde{f}+\ln\mu \tilde{R}} +3e^{2}f_{B} },\nonumber \\
&&\beta_{c}\frac{\partial \Gamma^{(2,0)}}{\partial c} = \beta_{c}\PR{e^{2} \PC{\frac{\partial \tilde{f}}{\partial c}+\ln\mu \frac{\partial \tilde{R}}{\partial c}}+ e^{3}\frac{\partial f_{B}}{\partial c}  },\nonumber\\
&& \beta_{v_{F}}\frac{\partial \Gamma^{(2,0)}}{\partial v_{F}}=  \beta_{v_{F}}\PR{e^{2} \PC{\frac{\partial \tilde{f}}{\partial v_{F}}+\ln\mu \frac{\partial \tilde{R}}{\partial v_{F}}}+ e^{3}\frac{\partial f_{B}}{\partial v_{F}}  }\nonumber \\
&& -i\beta_{v_{F}}\gamma^{i}p_{i}, \nonumber
\end{eqnarray}
where $\tilde{f}$ and $\tilde{R}$ stand for ${\rm finite}^{(2,0)}$ and ${\rm Res}^{(2,0)}$, respectively, and $f(B)=f_{B}$. Hence, Eq.~(\ref{RGE}) becomes
\begin{eqnarray}
&& e^{2}\tilde{R} +\beta_{e} \PR{2e\PC{\tilde{f}+\ln\mu \tilde{R}} +3e^{2}f_{B}}+ \beta_{c}e^{3}\partial_{c}f_{B} \nonumber \\
&&  +\beta_{c}e^{2} \PC{\partial_{c} \tilde{f}+\ln\mu \partial_{c} \tilde{R}} +\beta_{v_{F}}e^{2} \PC{\partial_{v_{F}} \tilde{f}+\ln\mu \partial_{v_{F}} \tilde{R}}\nonumber \\
&& + \beta_{v_{F}}e^{3}\partial_{v_{F}} f_{B} -i\beta_{v_{F}}\gamma^{l}p_{l}  -2\gamma_{\psi} \left[ -i\PC{\gamma^{0}p_{0}+v_{F}\gamma^{l}p_{l}} \right.\nonumber\\
&&\left. +e^{2}\PC{\tilde{f}+\ln\mu\tilde{R}}+e^{3}f_{B} \right] = 0,
\end{eqnarray}
where $\partial_{j}$ is a partial derivative with respect of one of the parameters $j=c,v_{F},e$.
We expand each of the $\beta_{j}$-functions and $\gamma_{\psi}$ in terms of $e$ up to third-order, e.g.,
\begin{eqnarray} 
\beta_{v_{F}}&=& \beta_{v_{F}}^{(1)} e +  \beta_{v_{F}}^{(2)} e^{2}+  \beta_{v_{F}}^{(3)} e^{3} +\ldots\ , \nonumber
\end{eqnarray}
and we unite the elements that share the same dependence on the coupling constant $e$. In this manner, we obtain three equations, one for each different order in $e$.

\textbf{a. Order of $e$}
\begin{eqnarray}
-i\gamma^{l}p_{l}\beta_{v_{F}}^{(1)} +2i\gamma^{(1)}_{\psi}(\gamma^{0}p_{0}+v_{F}\gamma^{l}p_{l}) = 0,\nonumber\\   \therefore \qquad  \gamma_{\psi}^{(1)}=0,\ {\rm and}\quad \beta_{v_{F}}^{(1)}=0.\nonumber 
\end{eqnarray}

\textbf{b. Order of $e^{2}$}
\begin{eqnarray}
&&\tilde{R}-i\gamma^{l}p_{l}\beta_{v_{F}}^{(2)} +2i\gamma^{(2)}_{\psi}(\gamma^{0}p_{0}+v_{F}\gamma^{l}p_{l}) = 0,\nonumber\\
&&\gamma^{l}p_{l}\PR{A_{2}-i\beta_{v_{F}}^{(2)}+ 2iv_{F}\gamma_{\psi}^{(2)}} + \gamma^{0}p_{0}\PR{A_{1}+2i\gamma_{\psi}^{(2)}}=0,\nonumber\\
&& \therefore \qquad \beta_{v_{F}}^{(2)} = -i\PC{A_{2}-v_{F}A_{1}}\quad {\rm and} \quad \gamma_{\psi}^{(2)}= \frac{i}{2}A_{1}.\nonumber
\end{eqnarray}
Here, we replaced $\tilde{R}$ as in Eq.~(\ref{Residue}), and we used that $\beta_{e}^{(1)}=0$, which can be obtained by doing the same procedure for the other two $\Gamma$-functions, i.e, $\Gamma^{(0,2)}$ and $\Gamma^{(2,1)}$. Note that $\tilde{R}$ only contains the divergent part of the electron self-energy. In other words, it is sufficient to compute $\Sigma^{(0)}$ to find $\beta_{v_{F}}^{(2)}$, which is precisely the function associated to the renormalization of the Fermi velocity. This is a second-order effect in the coupling constant $e$, and the magnetic field neither adds an extra term, nor changes this renormalization. Moreover, within the renormalization-group scheme seen in Eq.~(\ref{RGE0}), no finite contributions are encountered in this renormalization.

\textbf{c. Order of $e^{3}$}
\begin{eqnarray}
2\beta_{e}^{(2)}\PC{\tilde{f}+\ln\mu\tilde{R}}-i\beta_{v_{F}}^{(3)} \gamma^{l}p_{l} +2i \gamma_{\psi}^{(3)}  \PC{\gamma^{0}p_{0}+v_{F}\gamma^{l}p_{l}}=0,\nonumber
\end{eqnarray}
where we used the results $\beta_{j}^{(1)}=\gamma_{\psi}^{(1)}=0$. The magnetic field finite contribution would only be possible if $\beta_{e}^{(1)}\neq 0$. However, as the polarization tensor is finite in one-loop order, using dimensional regularization, its anomalous dimension is null, $\gamma_{A}^{(1)}=\gamma_{A}^{(2)}=0$, and this implies that both $\beta_{e}^{(1)}\ {\rm and}\ \beta_{e}^{(2)}$ are zero. Since $\gamma^{(3)}_{\psi}=0$, then $\beta^{(3)}_{v_{F}}=0$. Therefore, neither the linear magnetic field nor the other finite contributions change the Fermi-velocity renormalization.


\begin{thebibliography}{100}
\bibitem{footnote} In our opinion, the most appropriated use of the term weak magnetic field is to denote the regime of Shubnikov-de Haas oscillations, and not the regime of intermediate magnetic field, when several Landau levels are occupied. The strong magnetic field regime denotes the case when the physics involves only the lowest-Landau level.
\end{thebibliography}
\end{document}